Chapter #

# AST/RO
*A Small Submillimeter Telescope at the South Pole*

Antony A. Stark[*]

*Smithsonian Astrophysical Observatory; 60 Garden St.; Cambridge, MA 02138 USA*
aas@cfa.harvard.edu

Abstract:    Understanding of star formation in the Universe is advancing through submillimeter-wave observations of the Milky Way and other galaxies. Technological constraints on such observations require a mixture of telescope sizes and observational techniques. For some purposes, small submillimeter-wave telescopes are more sensitive than large ones. The Antarctic Submillimeter Telescope and Remote Observatory (AST/RO) is a small, wide-field instrument located at an excellent observatory site. By observing the Milky Way and Magellanic Clouds at arcminute resolution, it provides a context for interpreting observations of distant galaxies made by large interferometric telescopes. AST/RO also provides hands-on training in submillimeter technology and allows testing of novel detector systems.

## 1.      INTRODUCTION

The Antarctic Submillimeter Telescope and Remote Observatory (AST/RO, see Figure 1) is a 1.7-meter diameter submillimeter-wave telescope located at Amundsen-Scott South Pole Station. It is a small telescope operating in a wavelength band where other telescopes are much larger, a dwarf among giants and multi-headed behemoths. We shall see, however, that small, medium, and large submillimeter-wave telescopes all have particular roles to play in astronomical studies because of the nature of submillimeter-wave radiation and the technology used to detect it.

---

[*] Also at the Center for Astrophysical Research in Antarctica

The submillimeter is one of the most newly-developed and least explored bands in astronomy, at frequencies from about 300 Gigahertz up into the Terahertz range. Submillimeter-wave radiation is emitted by dense gas and dust between the stars, and submillimeter-wave observations allow us to study in unprecedented detail the galactic forces acting upon that gas and the star formation processes within it. Star formation is now visible throughout the Universe, from the earliest times to the present. We can observe submillimeter radiation in our own Galaxy and compare it to submillimeter radiation from other galaxies near and far. This chapter will discuss small, medium, and large submillimeter-wave telescopes, their relative strengths and sensitivities, and how they contribute to this great scientific undertaking.

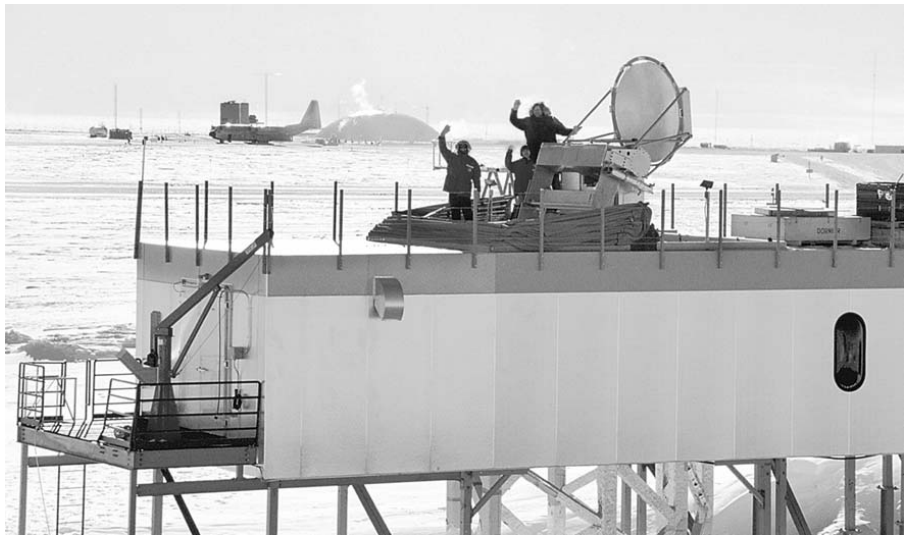

*Figure 1.* AST/RO at South Pole. The Antarctic Submillimeter Telescope and Remote Observatory atop its building at the South Pole in February 1997. The main part of the Amundsen-Scott South Pole Station lies beneath the dome, which is about 1 km distant. An LC130 cargo aircraft is parked on the skiway. (Photo credit: A. Lane)

*Table 1.* Bolometer Array Instruments.

| Instrument | Current Status | Number of Detectors, n |
|---|---|---|
| SCUBA[a] | operational | 128 |
| BOLOCAM[b] | in development | 151 |
| SAFIRE[c] | in development | 2,048 |
| SPECS[d] | proposed | 60,000 |

[a]http://www.jach.hawaii.edu/JACpublic/JCMT/scuba
[b]http://binizaa.inaoep.mx/pub/ins/pres_cam
[c]http://pioneer.gsfc.nasa.gov/public/safire
[d]http://www.gsfc.nasa.gov/astro/specs

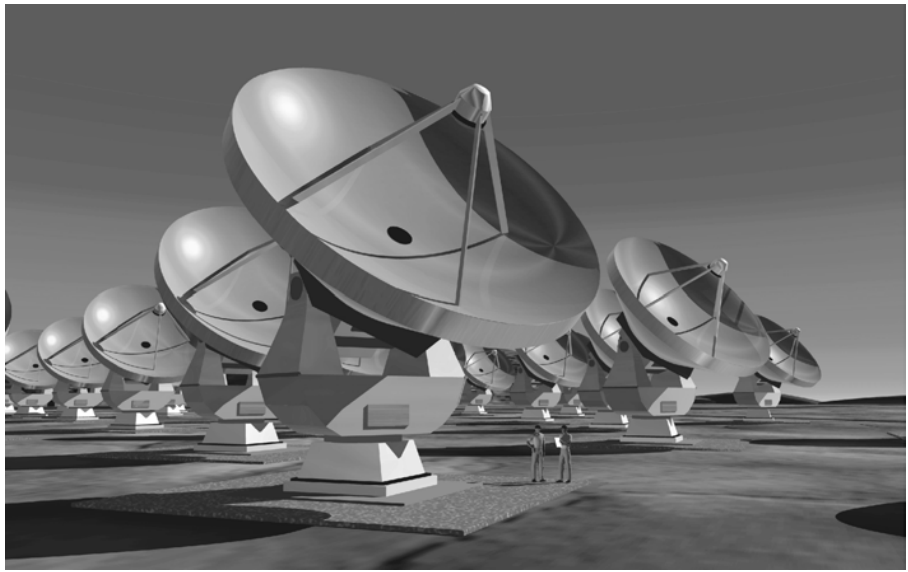

*Figure 2.* Conceptual design of the Atacama Large Millimeter Array (ALMA), a millimeter and submillimeter-wavelength interferometric telescope composed of 64 12-meter diameter antennas, to be located in the Chajnantor region of the Chilean Andes. (Image courtesy of ESO/NRAO/AUI)

## 2.        SUBMILLIMETER INSTRUMENTATION

The choice of what submillimeter-wave telescope to use for which observations depends on the instrumental trade-offs inherent in submillimeter-wave technology. This section surveys that technology. Three hypothetical observing scenarios illustrate the effect of telescope size on observations.

## 2.1     Heterodyne vs. Bolometer

Submillimeter waves lie between the radio and the infrared, and submillimeter-wave detectors can be either coherent detectors, as in radio astronomy, or incoherent detectors, as in infrared astronomy. In recent years both technologies have been dramatically improved (Carlstrom and Zmuidzinas 1995) and detectors are now at or near the "background limit", the point at which the detector noise is smaller than the noise from the atmosphere and telescope.

The coherent detectors are heterodyne mixers, where the signal from space is combined with a much stronger monochromatic "local oscillator" signal and then "mixed down" in a semiconductor diode or superconducting junction to produce a signal at an "intermediate frequency" (around 3 GHz), which is then processed using standard electronic techniques. Unlike radio astronomy at lower frequencies, the signal from space is not amplified prior to mixing, because there are as yet no practical submillimeter-wave low-noise amplifiers. The advantage of heterodyne receivers is that the intermediate frequency signal preserves all of the information content of the original submillimeter-wave signal. This means that spectroscopic analysis of the intermediate frequency signal by electronic means is the same as doing spectroscopy of the submillimeter-wave signal—very high resolution can easily be obtained. Also, the intermediate frequency signals from two or more antennas can be combined to make an interferometer, since the mixing process preserves the amplitude and phase of the submillimeter-wave signal. The disadvantages of heterodyne receivers are their cost and their limited bandwidth. At present, a single heterodyne detector plus its ancillary electronics costs at least a hundred thousand dollars. This cost is declining with time along with the costs of all electronics, but at present there are no heterodyne submillimeter-wave detector systems having more than a few mixers (Walker et al. 2001). The instantaneous bandwidth of submillimeter-wave heterodyne detectors is at most several Gigahertz, and this is a limitation on their sensitivity.

The incoherent detectors are bolometers, where the signal from space heats a tiny cryogenic thermometer and the resulting temperature rise is detected by a low-frequency electronic circuit. The advantages of bolometer detectors are their simplicity and their large bandwidths. It is becoming possible to manufacture bolometers and their ancillary electronics into integrated modules (Shafer et al. 2001) containing thousands of bolometers. The bolometers themselves have bandwidths hundreds of Gigahertz wide; the bandpass of bolometer detectors is limited by quasi-optical filters placed between the detector and the telescope. For spectroscopic applications this is a disadvantage, because it is not possible to make a truly narrowband quasi-optical filter for submillimeterwaves, so a spectrometer which makes use of bolometer detectors can resolve frequencies no better than 1 part in 5000. For broadband photometric applications, the huge instantaneous bandwidth of bolometers can make them more sensitive than heterodyne detectors. Since the signal from space is detected within the bolometer, the phase information in the signal is destroyed and bolometer detectors cannot be combined into phased interferometric arrays having many array elements. The emerging technology of large bolometer arrays (see Table 1) makes for fast, wide-field, inexpensive, sensitive detectors, but these can only be used on single-dish telescopes.

## 2.2    Interferometer vs. Single Dish

Both submillimeter-wave single-dish telescopes and the antenna elements of submillimeter-wave arrays are constructed like high-accuracy radio telescopes. The telescope mirrors consist of metal or carbon-fiber composite mirror panels supported by a truss of metal or carbon-fiber rods. Conventional fabrication techniques, applied with care, are capable of achieving structures more than 10 meters across, which have an overall accuracy of about 30 microns.

Submillimeter-wave interferometric arrays use essentially the same technology as do lower-frequency radio arrays like National Radio Astronomy Observatory's Very Large Array (VLA) or the Berkeley-Illinois-Maryland Array (BIMA). Submillimeter-wave heterodyne receivers are used at the front end of each array element. The intermediate frequency bands are collected and correlated using standard radio astronomy techniques (Thompson et al. 1986). Because submillimeter waves are short, all parts of the telescope system prior to the detection of the signal in the correlator must be constructed with care, so that the phase relationships between the intermediate frequency signals are maintained. It is important that neither the physical separation nor electrical pathlength between array elements vary by more than a few microns.

Submillimeter-wave telescopes operate in the diffraction limit. This means the resolution achievable by an array depends on the number of wavelengths between array elements, and the instantaneous field of view of an array depends on the number of wavelengths across a single array element. Large arrays have small synthesized beams, and arrays with large antennas as array elements have small fields of view. Since submillimeter waves are small, large submillimeter-wave interferometers have high resolution and small fields-of-view. The resolution and field-of-view of the Smithsonian Submillimeter Array (SMA, Moran 1998) and the Atacama Large Millimeter Array (ALMA, see Figure 2; Wild et al. 2000, Wootten 2001) are listed in Table 2.

The resolution achievable by a single-dish telescope depends on the number of wavelengths across the primary mirror, and the instantaneous field of view depends on the number of detectors filling the focal plane. The PoleSTAR array on AST/RO, for example, simultaneously feeds four independent heterodyne receivers, each observing in a slightly different direction. Each of the four beams is one arcminute in diameter. In a given time, four receivers observe four times as much sky, and obtain four times as many spectra, as would one receiver.

## 2.3    Large vs. Medium vs. Small

A small submillimeter-wave telescope can in some applications be more sensitive than a large submillimeter-wave telescope. Sensitivity in a telescope system can be expressed as the speed at which a particular observation can be made. The sensitivity of a submillimeter-wave telescope always depends on the noise in the detectors and the opacity of the atmosphere, but it does not always depend on the size of the telescope. In the case of a single pointing where the astronomical source fills the beam of the telescope, the sensitivity of large and small telescopes is the same, other things being equal. If the astronomical object is larger than the beam and a map of the object is desired, then the speed at which the observations proceed is proportional to the instantaneous field of view; this may favor the small telescope, which, other things being equal, will have the larger field of view. Consider now three idealized observing situations, one that favors a small single-dish telescope, one that favors a large single-dish telescope, and one that favors a large array instrument.

### 2.3.1    A large, uniform interstellar cloud.

This observation favors the small single dish. Suppose an interstellar cloud is 600 seconds of arc in diameter, and is approximately uniform in brightness within that diameter. This is typical of molecular clouds in our own Galaxy. We wish to measure the spatial and velocity extent of the $J = 4 \rightarrow 3$ CO line over the entire cloud. We observe the cloud with a 1-meter diameter single-dish telescope, a 10-meter diameter single-dish telescope, and an interferometric array consisting of sixteen 10-meter diameter dishes. The single-dish telescopes are each equipped with one heterodyne receiver with 100 K noise temperature, and the array is equipped with sixteen such receivers, one per array element. All telescopes are at the same site and have the same weather. To start the observations, we point each instrument at the center of the cloud and observe for one minute. The spectrometer on the 1-meter dish shows a line of 1 K brightness temperature in channels 1 km s$^{-1}$ wide, with a signal-to-noise ratio of 10. The spectrometer on the 10-meter dish shows an essentially identical line of 1 K brightness temperature in 1 km s$^{-1}$ channels, also with a signal-to-noise ratio of 10. The interferometric array shows…nothing. To the extent that the structure in the cloud is uniform over the 20 arcsecond field of view of the interferometer, the correlated power will be zero. In a real cloud, there would be some structure within the field of view and that structure might be detectable. In a practical array instrument, each element of the array would be simultaneously correlated with itself as well as with all the other elements. Each self-correlated signal would result in exactly the same spectrum as the single 10-meter dish, and the data from the ensemble of sixteen antennas would have a signal-to-noise ratio of 40, four times better than the single 1-meter dish or single 10-meter dish.

Now we decide to make a fully-sampled map of the cloud, with a uniform noise level of 0.1 K in 1 km s$^{-1}$ spectrometer channels. The 1-meter dish has a beamsize of 200 seconds of arc; a $10 \times 10$ grid of sampled positions covers the cloud, and the entire project is finished in 100 minutes, less than two hours. The 10 meter dish has a beamsize of 20 seconds of arc; a $100 \times 100$ grid of positions is needed to cover the cloud, and the project requires 10,000 minutes, about one week. This high-resolution map will show additional structure in the cloud, if it exists. The interferometric array can make the identical $100 \times 100$ map sixteen times faster, in about 10 hours. This map will show additional structure on scales of a few seconds of arc.

Because submillimeter radiation originates in the interstellar medium, it fills the spaces between the stars. Detectable submillimeter radiation from the Milky Way and the Magellanic Clouds covers much of the sky and is best studied by small telescopes, whereas submillimeter radiation from other galaxies is confined to regions of a few arcminutes or less and can only be seen by telescopes with large apertures.

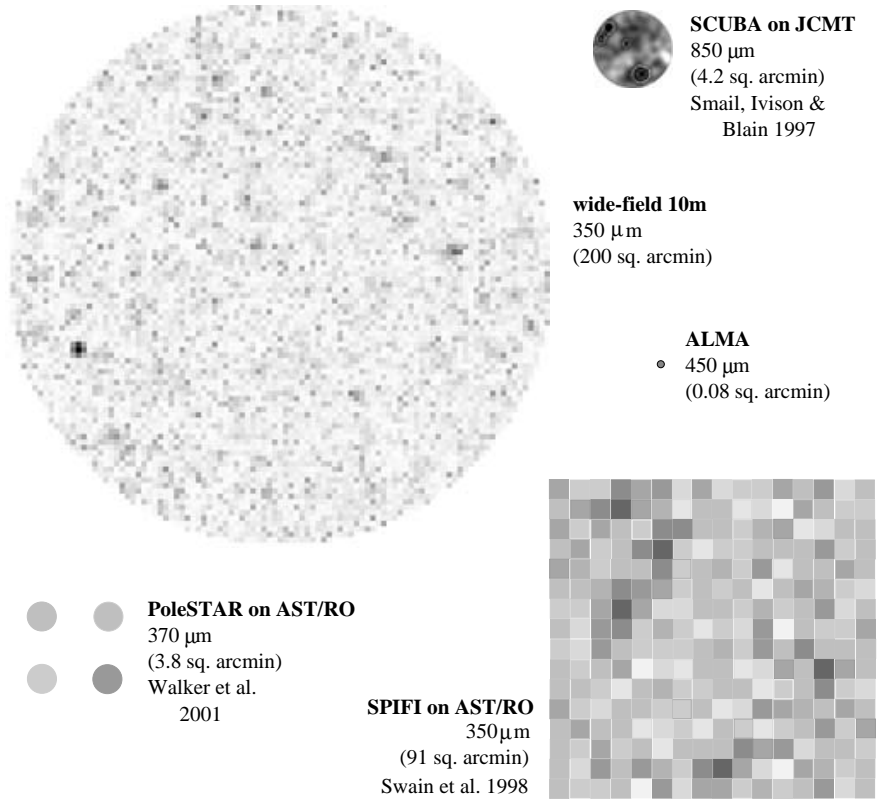

*Figure 3.* The field of view of submillimeter-wave telescopes. Simulated images from AST/RO and a hypothetical wide-field single-dish 10-m telescope are compared to actual SCUBA observations (Smail et al., 1997) and the image size of the ALMA. All images are to the same scale. The ALMA field contains a detailed spectroscopic image with thousands of pixels, too fine to be seen on this scale. PoleSTAR is a four-element heterodyne array currently operational on AST/RO, and SPIFI is an array spectrometer for AST/RO currently under development.

### 2.3.2 Search for protogalaxies.

This observation favors the large single dish. We search the sky for protogalaxies: these sources are about 2 seconds of arc in diameter and have a continuum brightness about 50 mJy at 350µm wavelength. Their positions on the sky are unknown. Both the 1-meter dish and the 10-meter dish are equipped with $16 \times 16$ focal-plane arrays of bolometers, like the SPIFI array shown in Figure 3. Each bolometer has an effective noise temperature of 100 K and a bandwidth $B = 100$ GHz. The array of sixteen 10-meter antennas are equipped with heterodyne receivers having an effective noise temperature of 100 K and a bandwidth $B = 10$ GHz. To start the observations, we point each telescope at a known, point-like continuum source and integrate for one minute. The noise in the observations is given by the radiometer equation:

$$F_{\text{noise}} = \frac{2kT_{\text{sys}}^*}{A\sqrt{Bt}}$$

where $T_{\text{sys}}^*$ is the effective atmosphere-corrected system temperature, $A$ is the total effective collecting area of the telescope, and $t \; (= 60 \text{ s})$ is the duration of the observation. Since all the detectors have the same noise temperature, and all the telescopes have the same sky conditions, they all have the same $T_{\text{sys}}^*$, which for definiteness we take to be 500 K. Since the source is point-like, it is unresolved in all the telescope beams and the sensitivity of the observations does depend on the collecting area of the telescope. For the 1-meter dish, $A = 0.5 \, \text{m}^2$; for the 10-meter dish, $A = 50 \, \text{m}^2$; and for the array, $A = 800 \, \text{m}^2$. The noise level after one minute is 1100 mJy for the 1-meter dish, 11 mJy for the 10-meter dish, and 2.2 mJy for the array. To achieve a noise level of 10 mJy in a single pointing requires 8.4 days for the 1-meter dish, 1.2 minutes for the 10-meter dish, and 3 seconds for the interferometric array.

Now we want to search a square degree of sky for protogalactic sources, at a survey noise level of 10 mJy. Observing with the 1-meter dish, it requires 15 pointings of the $16 \times 16$ array to cover one square degree (see Figure 3), and the entire survey can be accomplished in 126 days. The focal-plane array on the single 10-meter dish requires 1500 pointings, and the survey is accomplished in 1.2 days. The interferometric array has a field of view only 20 arcseconds across, and must be repointed $\sim 4 \times 10^5$ times to cover a square degree; even at 3 seconds per point the survey requires two weeks.

The speed with which a survey by a telescope system can detect unknown point sources of radiation is proportional to

$$\text{figure of merit} \equiv \frac{nBA}{T_{\text{sys}}^{*\,2}},$$

where $n$ is the total number of detectors or receivers. This holds regardless of whether the detectors are bolometers or heterodyne mixers, the way the total collecting area $A$ is distributed across antennas, or the way the detectors are arranged among antennas, provided each detector is coupled to a diffraction-limited beam (Stark 2000). The "figure of merit" is proportional to $A$ and not $A^2$ because the area of sky covered by the field of view varies as $A^{-1}$. In a contest to maximize this figure of merit, single-dish instruments have a big advantage because bolometer focal-plane arrays could have $n \sim 10^3$ or even $10^4$, and $B$ is intrinsically large. The number of detectors, $n$, is limited by the size of the telescope's field of view—the larger the diameter of the single dish, the more pixels can be crammed into the field of view (cf. Stark 2000). A future single-dish telescope tens of meters in diameter with tens of thousands of bolometers in a focal-plane array would be a powerful instrument. The expression for figure of merit also shows the importance of atmospheric opacity, $\tau$, since $T_{sys}^{*}$ increases with opacity even more quickly than $\exp(\tau)$. An increase in opacity from $\tau = 1$ to $\tau = 2$ slows observations by an order of magnitude.

Table 2 shows the continuum sensitivity and beamsize of actual submillimeter-wave telescopes, both proposed and operational, and illustrates the strength of wide-field single-dish telescopes for deep continuum surveys. The noise equivalent flux density (NEFD) values in this table for each of the instruments originate if possible from the scientific group operating or proposing that instrument; they are a mixed bag of real results on the telescope and calculations, some of which are based on optimistic assumptions.

For speed and coverage in deep surveys, a wide-field single-dish telescope can pursue an optimization strategy:

1.  Build a reflector whose beamsize is equal to the spatial scale of interest.
2.  Make the telescope losses, blockage, and spillover as small as possible at the frequencies of interest.
3.  Make the field of view as large as possible.
4.  Populate that field of view with as many broadband detectors as possible.
5.  Place the telescope at the best possible site.

Such a wide-field single dish is an "optimal design" for the discovery of objects at spatial frequencies near $\frac{1}{2}D/\lambda$.

It is a waste of scarce resources to search for protogalaxies with an interferometric array—the efficient and cost-effective observing strategy is to search using a focal-plane bolometer on a medium-sized single-dish telescope, and then study the detected objects in detail using an interferometer.

### 2.3.3    Observations of galaxies.

A resolution of one second of arc or better requires a telescope aperture greater than $72\,\text{m} \times (\lambda/350\,\mu\text{m})$, but the largest submillimeter-wave single dish is the 15 m diameter James Clerk Maxwell Telescope, and there are at present no plans for bigger single dishes. Resolution better than a few seconds of arc is therefore the sole domain of interferometers. As shown in Table 2, the sensitivity of the large arrays is unsurpassed for unresolved objects having known positions. Among astronomical objects, this includes much of the Universe: almost all galaxies, as well as regions of star formation within our own Galaxy. Large submillimeter-wave interferometers are tremendously powerful instruments, and they alone are capable of looking within distant galaxies and into the cores of star-forming clouds.

*Table 2.* Continuum Sensitivity of Submillimeter-wave Telescopes

| *Telescope* | $A^a$ (m²) | $R^b$ (″) | $S^c$ | $NEFD^d$ (mJy s½) 850μm | 350μm | *Time in hours to survey 1 square degree at 1 mJy* 850μm | 350μm |
|---|---|---|---|---|---|---|---|
| wide-field 10-m | 79 | 11 | 200 | 60 | 74 | 18 | 27 |
| wide-field 30-m | 711 | 4 | 22 | 7 | 8 | 2 | 3 |
| AST/RO | 2 | 65 | 92 | 2160 | 2660 | $5.1 \times 10^4$ | $7.7 \times 10^4$ |
| JCMT | 177 | 7 | *5* | *80* | 760 | *$1.3 \times 10^3$* | $1.2 \times 10^5$ |
| CSO | 79 | 11 | 11 | 150 | 2200 | $2.0 \times 10^3$ | $4.4 \times 10^5$ |
| SOFIA | 5 | 44 | 50 | | 200 | | 800 |
| Herschel | 7 | 32 | 4.3 | | 54 | | 583 |
| "submm CBI"ᵉ | 8.3 | 11 | 4 | 297 | 2529 | $6.2 \times 10^3$ | $1.1 \times 10^7$ |
| SMA | 227 | 2 | 0.14 | 134 | 1250 | $3.6 \times 10^4$ | $1.9 \times 10^7$ |
| MK arrayᶠ | 483 | 0.5 | 0.02 | 59 | 790 | $4.9 \times 10^4$ | $5.2 \times 10^7$ |
| ALMA | 7000 | 0.2 | 0.06 | 2 | 15 | 19 | $6.4 \times 10^3$ |

a. Telescope area (m²)

b. Resolution element (arcsec) for λ=450μm. Resolution element scales as λ.

c. Instantaneous sky coverage (arcmin²) for λ=450μm. Instantaneous sky coverage scales as λ² for interferometers, is roughly independent of λ for most single-dish instruments.

d. Noise Equivalent Flux Density, the sensitivity to point sources whose positions are known. Numbers in italics are measured, on-the-telescope values and are subject to downward revision with improved techniques. Predicted sensitivities are optimistic in the sense that in all cases they are near the thermal background limit, a limit that has not yet been achieved in practical submillimeter bolometer systems. This table is based on the work of Hughes and Dunlop, 1997.

e. A hypothetical submillimeter array with the configuration of the Cosmic Background Imager (Cartwright et al. 2001); the actual CBI operates at wavelengths near 1 cm.

f. Mauna Kea array consisting of the SMA, the CSO, and the JCMT.

## 3.    AST/RO AS A SMALL TELESCOPE

AST/RO is a 1.7-m diameter single-dish telescope, operating as a user-facility observatory (Stark et al. 2001). It has an offset optical design, which allows for the mounting of large receivers in a warm environment while providing a clean radio beam (Stark et al., 1997). AST/RO is located at the geographic South Pole, an excellent submillimeter-wave observatory site (Lane 1998; Chamberlin 2001). AST/RO's logistical needs are supplied through the U. S. National Science Foundation Amundsen-Scott South Pole Station. One advantage of AST/RO's small size is its relatively small power and dormitory requirements: it would not be possible to support ALMA at the Pole—this would require South Pole Station and the United States Antarctic Program to be several times larger than they are. AST/RO takes advantage of excellent site conditions that are not available to large interferometric arrays

## 3.1        AST/RO as Complement to Larger Telescopes

In its first six years of operation, AST/RO has concentrated on heterodyne spectroscopy. AST/RO has made position-position-velocity maps of submillimeter-wave spectral lines such as the $^3P_1 \rightarrow ^3P_0$ and $^3P_2 \rightarrow ^3P_1$ fine-structure lines of atomic carbon (C I) and the $J = 4 \rightarrow 3$ and $J = 7 \rightarrow 6$ lines of carbon monoxide (CO) with arcminute resolution over regions of sky which are several square degrees in size. AST/RO can observe molecular clouds throughout the fourth quadrant of the Milky Way and the Magellanic Clouds, in order to locate star-forming cores and study in detail the dynamics of dense gas in our own Galaxy. AST/RO studies of molecular clouds with varying heavy element content under a variety of galactic environments are showing how molecular clouds are structured, how the newly-formed stars react back on the cloud, and how galactic forces affect cloud structure. AST/RO studies of the Galactic Center region have shown, for example, that the smooth ring of molecular material 300 parsecs from the Galactic Center is on the verge of coagulating into a single, massive molecular cloud like the one surrounding Sgr B2, and will likely undergo a burst of star formation in a few hundred million years (Kim et al. 2000). AST/RO studies have shown that molecular clouds structure is affected by their heavy element content (Bolatto et al. 2000) and by their proximity to spiral arms (Zhang et al. 2001). AST/RO has observed the isotopically-shifted $^{13}$C I line, in order to study the gradient of heavy elements in the Galaxy (Tieftrunk et al. 2001).

AST/RO's primary beam size is $66'' \times (\lambda/350\mu\text{m})$, about a minute of arc. This is similar to the resolution achieved by the Arecibo telescope and the NRAO 12 meter telescope at their operating wavelengths. It is a good scale at which to study the interstellar medium in the Milky Way. High resolution is always desirable, but as we have seen, technology places limits on the speed at which the sky can be mapped with high resolution. To study the Milky Way as a galaxy, it is essential to observe hundreds or even thousands of square degrees of sky. AST/RO was designed as a practical compromise, an instrument with sufficient resolution to study interstellar structure but cheap and fast enough to study the Milky Way as a whole. AST/RO does for the Milky Way what the SMA and ALMA will do for other galaxies. It is essential to first understand our own Galaxy, in order to interpret the observations of other galaxies.

## 3.2      AST/RO as Platform for Instrumental Prototypes

Two new short wavelength instruments for AST/RO are in development:

- Dr. G. Stacy and collaborators have developed the South Pole Imaging Fabry-Perot Interferometer (SPIFI, Swain et al., 1998), a 25-element bolometer array preceded by a tunable Fabry-Perot filter. This instrument was successfully used on the JCMT in May 1999 and April 2001 and is being modified with new instrumentation, cryogenics, and detectors for South Pole use. SPIFI will be upgraded to a $16 \times 16$ array, as shown in Figure 3. SPIFI is frequency agile and can observe many beams at once, but has limited frequency resolution ($\sim 100$ km s$^{-1}$) and scans a single filter to build up a spectrum.

- Dr. S. Yngvesson and collaborators (Gerecht et al. 1999; Yngvesson et al. 2001) are developing a 1.5 THz heterodyne receiver, the Terahertz Receiver with Niobium Nitride Device (TREND) for installation on AST/RO in 2002. Preliminary lab tests indicate TREND will be the most sensitive receiver ever to operate in this important frequency band. It will have a single pixel and utilize a state-of-the-art, turn-key, laser local oscillator. TREND will initially observe three different spectral lines by retuning its laser or changing the gas used. Although the HEB mixer could in principle be operated with solid-state multiplier local oscillator sources, the laser provides a more flexible and reliable local oscillator injection scheme and allows AST/RO to expand in the future to THz multi-pixel receivers. The frequency resolution of TREND is high ($\sim 0.4$ km s$^{-1}$), limited by the stability of the laser.

In addition, Dr. D. Prober and collaborators have proposed a low-noise, low power 1.5 THz heterodyne receiver based on aluminum and tantalum HEB technology which they will test on AST/RO.

South Pole site testing suggests that the 200 μm wavelength atmospheric window has a median winter transparency of order 10%. All previous observations at these wavelengths have been done from airborne or spaceborne platforms; a transparency of 10% is good enough to allow more extensive, higher resolution observations to be done from the ground. The atomic and molecular transitions accessible in this spectral range are important diagnostics of a variety of astronomical phenomena including planetary atmospheres, star-forming regions, young stellar objects, circumstellar envelopes, planetary nebulae, starburst galaxies, and molecular clouds. Deployment of these technologies on a ground-based telescope is a path of technological development that has exciting prospects. On AST/RO, the $\sim 35''$ beam size and high spectral resolution ($\sim 0.4$ km s$^{-1}$) of Terahertz receivers will allow the study of galactic star-forming regions and large-scale studies of nearby galaxies. In the future, these detectors could be used on the South Pole Submillimeter Telescope (SPST), an 8-meter telescope (NRC 2001), which would have a beamsize of $\sim 7''$.

## 3.3     AST/RO as an Educational Tool

During the Austral winter observing season, when South Pole Station is closed to incoming or outgoing transport for a period of 8 months, AST/RO is attended by a Winterover Scientist in residence at the Pole. The AST/RO Winterover Scientist is a three-year research and training position at the Postdoctoral level. The first year includes an Austral summer season at the telescope, learning its systems and participating in engineering and maintenance work. Hands-on experience is gained in all techniques of submillimeter-wave observatory work: electronics, cryogenics, detectors, vacuum technology, data acquisition and reduction software. The Winterover Scientist while in residence at the Smithsonian Astrophysical Observatory (SAO) in Cambridge, Massachusetts, develops research plans for AST/RO. The second year is the winterover, and the AST/RO Winterover Scientist is responsible for all on-site aspects of observatory operations. Winterover Scientists carry out their own proposed projects in addition to all other observations. The third year is again spent in residence at SAO, preparing data for publication and helping to train new Winterover Scientists. The intensive instrumental and observing experience afforded by AST/RO provides valuable hands-on training with state-of-the-art equipment. AST/RO serves as a training instrument for the submillimeter-wave astronomers of the future.

## 4.    SUMMARY

AST/RO is a small submillimeter-wave telescope. Its small size and modest support requirements have allowed it to be deployed to the South Pole, an excellent submillimeter site with unique logistical challenges. AST/RO is studying submillimeter-wave radiation throughout the Milky Way and Magellanic Clouds. AST/RO observations do for our own Galaxy what the large interferometric arrays will do for other galaxies. AST/RO also provides a platform for hands-on training in submillimeter-wave technology and for testing of new instruments.

In submillimeter-wave astronomy, small and medium single-dish telescopes as well as giant array instruments all have a role. Submillimeter-wave detector technology and the atmosphere impose limits on the speed at which observations can be performed. Observations on a large scale call for single-dish telescopes with large arrays of focal-plane detectors; observations on a small scale require interferometric arrays. In the coming decades, submillimeter-wave observations with interferometric arrays will bring new understanding of star formation processes throughout the Universe, from the earliest times to the present. It is important to recognize, however, that these giant arrays need "finder telescopes", single-dish telescopes equipped with focal-plane arrays that will survey the submillimeter sky on large scales. Submillimeter astronomy needs Schmidt camera equivalents in addition to giant glass equivalents. Observations of the Milky Way with small telescopes will provide the context for understanding interferometric observations of other galaxies. Single-dish surveys of the sky with large bolometer arrays will discover observing targets for the interferometric arrays that are not detectable at other wavelengths.

## Acknowledgments

This work was supported in part by the Center for Astrophysical Research in Antarctica and the National Science Foundation under Cooperative Agreement OPP89-20223. I thank Adair P. Lane and Wilfred Walsh for their comments on the manuscript.

Filename:           mychapter8a.doc
Directory:          C:\WINDOWS\Desktop\smalltelescopes
Template:           C:\WINDOWS\Profiles\aas\Application
     Data\Microsoft\Templates\vbaKAPedvo.dot
Title:              Chapter
Subject:
Author:             Antony A. Stark
Keywords:
Comments:
Creation Date:      10/16/2001 7:08 PM
Change Number:      5
Last Saved On:      10/18/2001 10:53 AM
Last Saved By:      Antony A. Stark
Total Editing Time: 39 Minutes
Last Printed On:    10/18/2001 11:52 AM
As of Last Complete Printing
     Number of Pages:   17
     Number of Words:   5,386 (approx.)
     Number of Characters:   30,703 (approx.)